\documentclass{article}
\usepackage{spconf,amsmath,graphicx}

\usepackage[absolute]{textpos}
\usepackage[font=small,skip=2pt]{caption}
\newcommand\blfootnote[1]{%
  \begingroup
  \renewcommand\thefootnote{}\footnote{#1}%
  \addtocounter{footnote}{-1}%
  \endgroup
}

\title{Automatic artifact removal of resting-state fMRI \\ with Deep Neural Networks}
\name{Christos Theodoropoulos$^{\star}$ \qquad Christos Chatzichristos$^{\dagger}$ \qquad Sabine  Van Huffel$^{\dagger}$}
\address{$^{\star}$ Department of Computer Science, LIIR Lab, KU Leuven, Belgium \\
         $^{\dagger}$ Department of Electrical Engineering, STADIUS, KU Leuven, Belgium}
\begin{document}
\begin{textblock}{6}(1.5,1)
\noindent Preprint. Under review.
\end{textblock}
%
\maketitle
\begin{abstract}

\blfootnote{\copyright 2021 IEEE. Personal use of this material is permitted. Permission from IEEE must be obtained for all other uses, in any current or future media, including reprinting/republishing this material for advertising or promotional purposes, creating new collective works, for resale or redistribution to servers or lists, or reuse of any copyrighted component of this work in other works.}

Functional Magnetic Resonance Imaging (fMRI) is a non-invasive technique for studying brain activity. During an fMRI session, the subject executes a set of tasks (task-related fMRI study) or no tasks (resting-state fMRI), and a sequence of 3-D brain images is obtained for further analysis. In the course of fMRI, some sources of activation are caused by noise and artifacts. The removal of these sources is essential before the analysis of the brain activations. Deep Neural Network (DNN) architectures can be used for denoising and artifact removal. The main advantage of DNN models is the automatic learning of abstract and meaningful features, given the raw data. This work presents advanced DNN architectures for noise and artifact classification, using both spatial and temporal information in resting-state fMRI sessions. The highest performance is achieved by a voting schema using information from all the domains, with an average accuracy of over 98\% and a very good balance between the metrics of sensitivity and specificity (98.5\% and 97.5\% respectively).
\end{abstract}
\section{Introduction}
\label{sec:intro}

Currently, one of the most widely used techniques for studying and analyzing brain connectivity and activity is fMRI. During an fMRI experiment, random noise and artifacts are introduced (e.g. heartbeat, head motion, thermal noise, etc). Moreover, the noise can be related to the specific hardware and the nature of the experiment. A successful and substantial analysis of the fMRI session requires high quality, noise-free data. Hence, the robust denoising and artifact removal is a crucial step of the fMRI processing \cite{2008_lindquist_statistical}. This task is challenging because some types of noise are difficult to be detected due to the fact that they are very rare or quite similar to regular components \cite{2015_bright_is}. 
\par

Blind Source Separation (BSS)~\cite{2015_theodoridis_machine} is a very important step for interpreting and analyzing the fMRI data. The localization of the activated brain areas is a challenging BSS task, in which the sources consist of a combination of spatial maps (areas activated) and time-courses (timings of activation) \cite{2019_chatzichristos_journal}. The sources should be classified, for clean-up purposes, as artifacts or neuronal signals. Both temporal and spatial information is used to categorize the source as noise/artifact or neuronal signal, the sources classified as artifacts are removed during the reconstruction of the signal. Independent Component Analysis (ICA) \cite{calhoun2009review} is a statistical method which tries to find a linear transformation of the observable space into a new space such that the individual new variables are mutually independent. ICA is a powerful technique for separating the various source of fluctuations and, ICA assumes that statistically independent spatial maps are mixed with the use of corresponding time-courses in an associated (mixing) matrix.
\par

The most widely used Machine Learning based approach for artifact removal is FIX ("FMRIB's ICA-based X-noiseifier") \cite{salimi2014automatic}, \cite{griffanti2014ica}. It is an ICA-based framework using FastICA algorithm (as implemented in Melodic toolbox \cite{beckmann2004probabilistic}). Principal Component Analysis (PCA) \cite{wold1987principal} is applied as a preprocessing step, for dimensionality reduction and reduction of unstructured noise. The features (over 180) are manually engineered in order to capture aspects of spatial maps (e.g. size of the clusters and voxels overlaying bright/dark raw data voxels) timeseries, and frequency spectrum (e.g. autoregressive and distributional properties, jump amplitudes). The hand-crafted features are sensitive to the acquisition and preprocessing parameters. Hence, the re-training of the model is essential when the data differ a lot from the initial data, which were used for the training of the models. Finally, multiple different classifiers are stacked in order to extract the final decision.
\par

In the view of the DNN success in various biomedical problems \cite{mao2019spatio}, \cite{duc20203d}, a Deep Learning \cite{goodfellow2016deep} framework is proposed for automatic noise and artifact detection in resting state fMRI data \cite{kam2019deep}, which exhibits good performance. The dataset of the study is taken from Baby Connectome Project (BCP \cite{howell2019unc}) and contains resting state sessions from 32 subjects/infants. ICA is applied on the data and 150 components per subject are extracted. Trained raters decided whether a component is related to noise or a nuisance signal. Normalization (standardization) is applied on each extracted 3D spatial map. The proposed framework contains a 3D Convolution Neural Network (CNN) \cite{lecun1995convolutional} model which receives the spatial maps as input and extracts meaningful spatial features. The temporal information is analyzed by a 1D CNN model, which learns high-level temporal representations. For each convolutional layer, ReLU is used as activation function. Finally, a stack of fully connected layers is added in each model, in order to perform the classification of the component (signal or noise/artifact). A majority voting schema is also applied for the final classification.
\par

In this study, advanced Deep Learning architectures are used for denoising and artifact removal. Having as starting point the proposed models of \cite{kam2019deep}, we want to explore the effectiveness of more complex architectures and the addition of frequency information as input. The labeled extracted components are used for training and evaluation of the Deep Learning models. After necessary data-preprocessing and manipulation, the training and testing of the different DNN models are executed. The performance is tested, using the spatial, temporal, and frequency information independently and jointly. The main outcome of the study is twofold; spectral information boosts the overall performance and a weighted voting schema achieves the best results.

\section{DEEP LEARNING METHODS}
\label{sec:DL models}

The proposed DNN models can be separated based on the given input (spatial, temporal, and frequency). The main layer of the models using 3D spatial maps as input is the convolutional layer, which is capable of extracting high-level feature representations taking into account the local connectivity between the elements of the input. We employ models using both temporal and frequency information in order to test whether the assumption used in \cite{kam2019deep}, that a neural network using temporal information can infer all the meaningful frequency features, is valid, and whether we can improve the total performance. 

\subsection{Models using spatial information}
 The first model (\(CNN_{sm_1}\)), shown in Fig. 1, is similar to the one proposed in \cite{kam2019deep} and is considered as the baseline model. The main difference is that the stride of every convolution operation is set to 1, while in \cite{kam2019deep} stride values of 2 and 3 are used. The second model (\(CNN_{sm_2}\), Fig. 2) has a slight difference with the first one. Batch Normalization (BN) \cite{ioffe2015batch} layers are used after each convolutional layer. BN layer \cite{liao2016importance}, \cite{tian2019enhanced} helps the network to get trained smoother and faster, decreases the sensitivity to the weight initialization and can be used as a type of regularization. Hence, the second model tests whether the addition of BN layer is advantageous in our task. The third model (\(CNN_{sm_3}\), Fig. 3) includes the idea of residual blocks. This type of block is initially proposed in ResNet architecture \cite{he2016deep} and contains skip connections, which help the network to learn additional residual features. Learning residual features boosts the performance in many computer vision tasks \cite{he2016deep}, \cite{szegedy2016inception}. Hence, we want to investigate whether residual blocks are efficient in our study. ReLU is used as the activation function in all of the layers (3D convolutional layers and fully connected layers) of the proposed models.

\begin{figure}[t]
  \centering
  \centerline{\includegraphics[width=9cm,height=5cm,keepaspectratio]{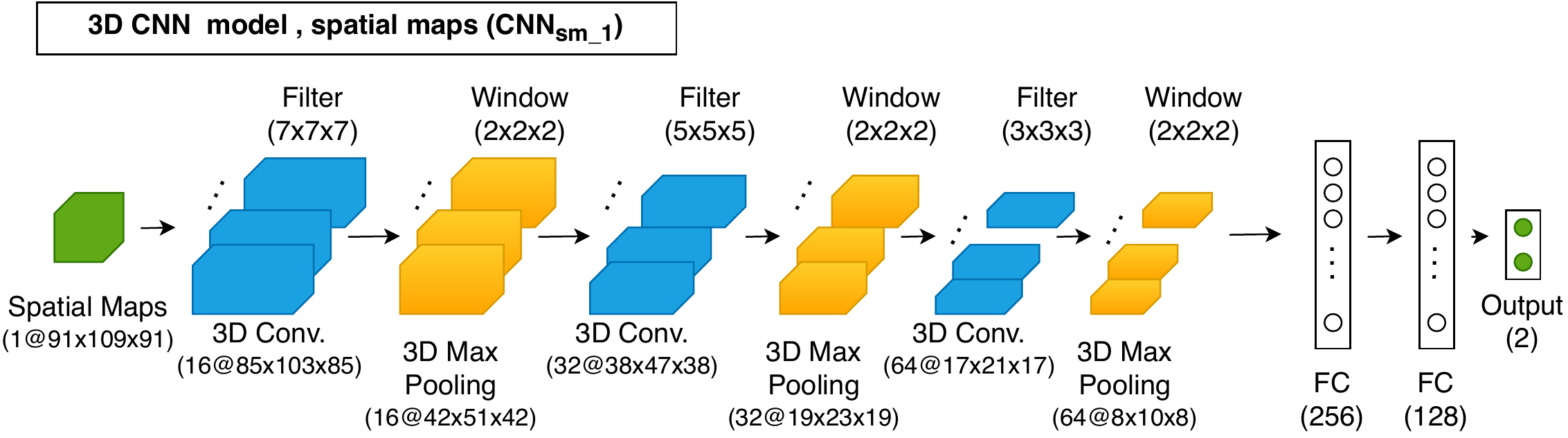}}
  \caption{\(CNN_{sm_1}\) model}
  \vspace{-2mm}
\end{figure}

\begin{figure}[!h]
  \centering
  \centerline{\includegraphics[width=9cm,height=5cm,keepaspectratio]{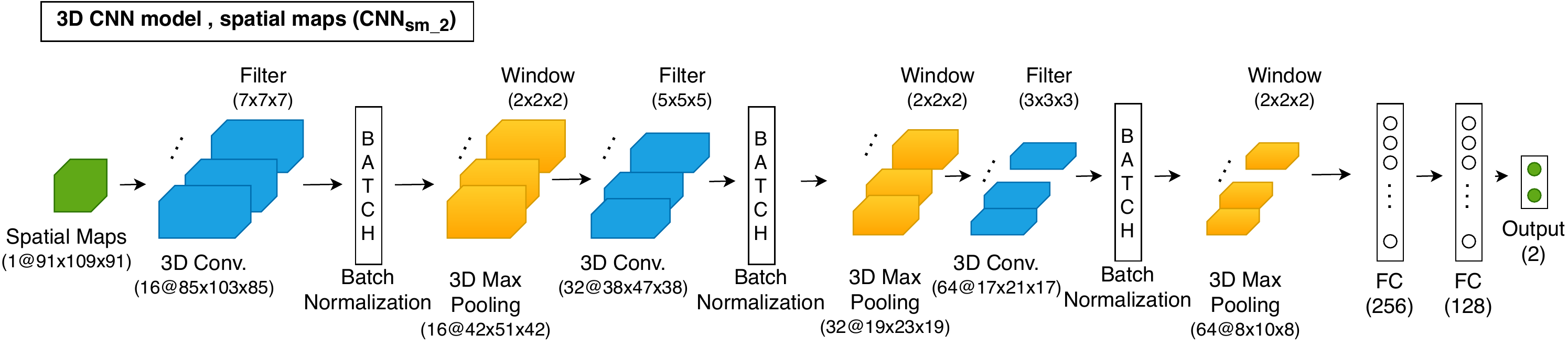}}
  \caption{\(CNN_{sm_2}\) model}
    \vspace{-2mm}
\end{figure}

\begin{figure}[!h]
  \centering
  \centerline{\includegraphics[width=9cm,height=5cm,keepaspectratio]{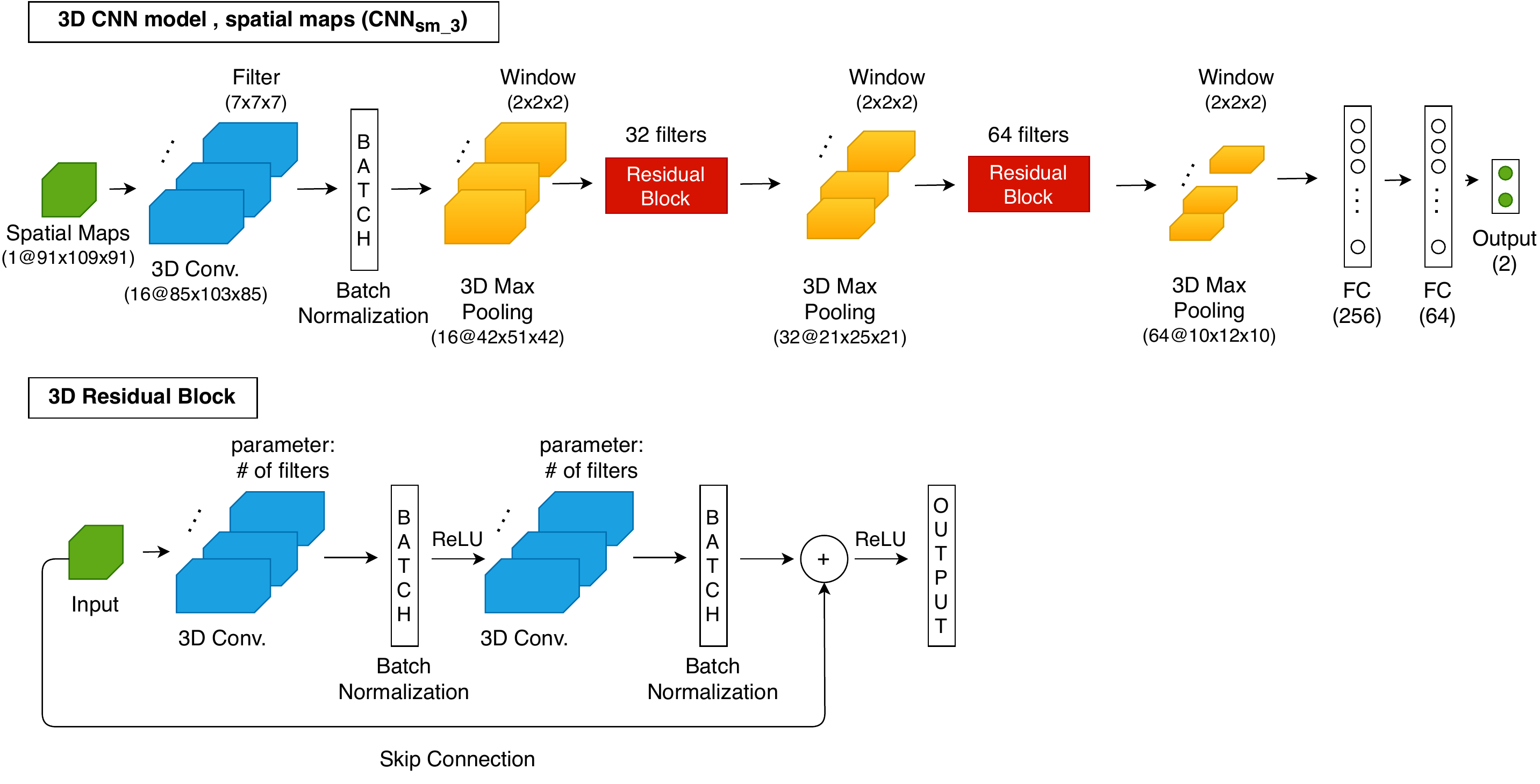}}
  \caption{\(CNN_{sm_3}\) model}
    \vspace{-2mm}
\end{figure}

\begin{figure}[!h]
  \centering
  \centerline{\includegraphics[width=9cm,height=5cm,keepaspectratio]{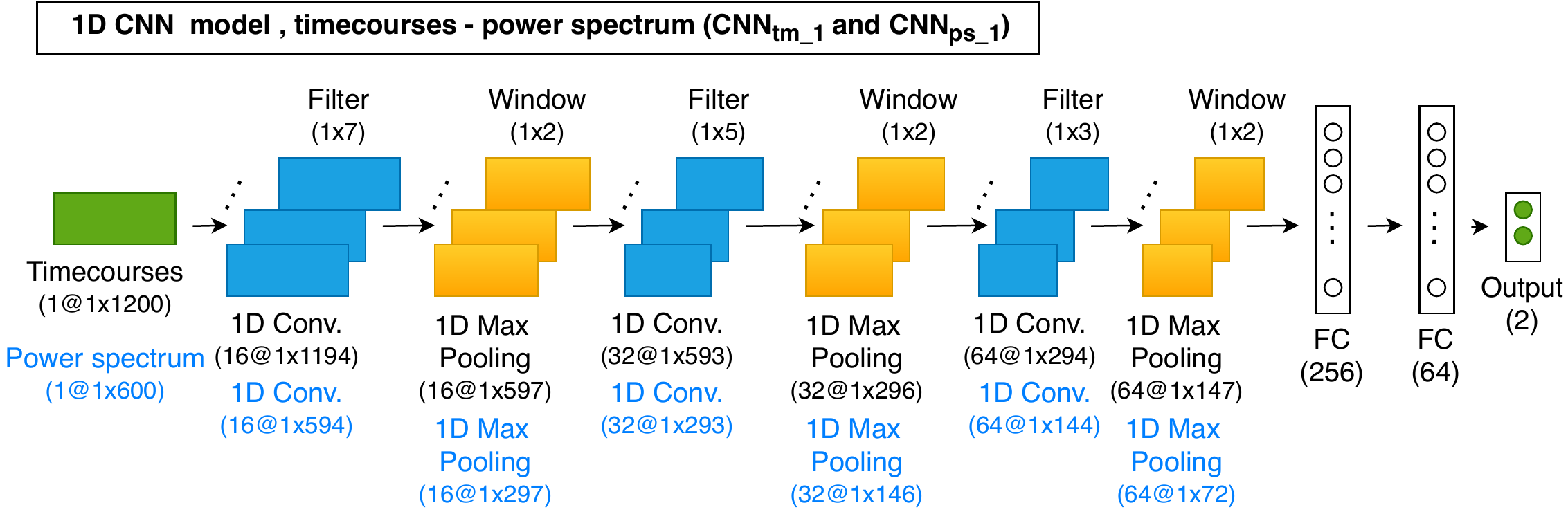}}
  \caption{\(CNN_{tm_1}\) and \(CNN_{ps_1}\) models}
    \vspace{-2mm}
\end{figure}

\begin{figure}[!h]
  \centering
  \centerline{\includegraphics[width=9cm,height=5cm,keepaspectratio]{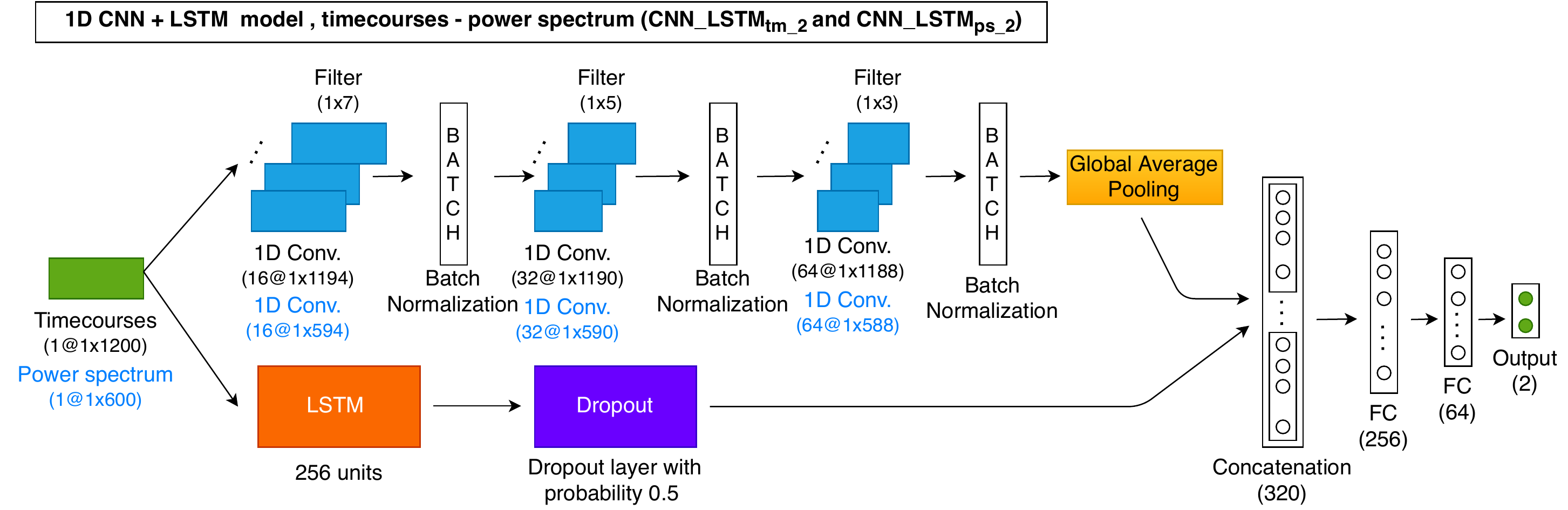}}
  \caption{\(CNN-LSTM_{tm_2}\) and \(CNN-LSTM_{ps_2}\) models}
    \vspace{-2mm}
\end{figure}

\noindent Only the last output layer uses the Sigmoid activation function in order to extract the final probability (1: perfect noise, 0: pure signal).

\subsection{Models using temporal and frequency information}
The proposed architectures using temporal and frequency information are identical. The first model (\(CNN_{tm_1}\), \(CNN_{ps_1}\), Fig. 4), which is used as baseline model, employs a sequence of 1D convolutional and max pooling layers. It is similar to the model proposed in \cite{kam2019deep}. The second model (\(CNN-LSTM_{tm_2}\), \(CNN-LSTM_{ps_2}\), Fig. 5) introduces a parallel architecture, which also includes an LSTM block \cite{hochreiter1997long}, followed by a dropout layer. The usage of LSTM block \cite{dvornek2018combining}, \cite{yan2019discriminating} provides the capability of learning long-term time-dependent patterns. The dropout layer is used for regularization in order to avoid overfitting.

\section{RESULTS}
\label{results}

The dataset consists of resting-state fMRI data of young healthy adults from Human Connectome Project \cite{smith2013resting}, \cite{van2013wu}. ICA is applied on the data for unmixing the different sources and each extracted component is labeled as noise/artifact or signal. The first step of the experimental process is the separation of the three different subsets of the dataset: training, validation, and test set. Taking into account the computational cost of the training of the models, 80 subjects are included in the training set and 20 subjects in the validation set. In the training set, a random sampling is performed for each different split (5-fold cross-validation technique) in order to balance the classes and handle the imbalance problem, as the class which contains the noisy components is dominant. The remaining 294 subjects are used as test set. Hence, as the number of subjects in the test set is large, the evaluation process indicates robustly the generalization capabilities of the models.
\par

As 5-fold cross-validation technique is used, the models are trained five times. For all the models, Adam \cite{kingma2014adam} is used as optimizer with learning rate equal to 0.001. For the models using spatial information (\(CNN_{sm_1}\), \(CNN_{sm_2}\), and \(CNN_{sm_3}\)) the batch size is set to 16 and early stopping is applied after 3 epochs, when no performance improvement is achieved in the validation set. For the models using temporal and frequency information (\(CNN_{tm_1}\), \(CNN-LSTM_{tm_2}\), \(CNN_{ps_1}\), and \(CNN-LSTM_{ps_2}\)) the batch size is set to 128 and early stopping is applied after 4 epochs. 
\par

Other than training the different models separately, we also train four combinations of them with the addition of a concatenation layer and two fully connected layers with 128 and 32 neurons, in order to check for a possible increment in the performance. The tested combinations are the following: 

\begin{itemize}
    \vspace{-2mm}
    \item \(Comb_{1}\): \(CNN_{sm_1}\), \(CNN_{tm_1}\) and \(CNN_{ps_1}\)
    \vspace{-2.5mm}
    \item \(Comb_{2}\): \(CNN_{tm_1}\) and \(CNN_{ps_1}\)
    \vspace{-2.5mm}
    \item \(Comb_{3}\): \(CNN_{sm_1}\) and \(CNN_{tm_1}\)
    \vspace{-2.5mm}
    \item \(Comb_{4}\): \(CNN-LSTM_{tm_2}\) and \(CNN-LSTM_{ps_2}\).
    \vspace{-5mm}
\end{itemize}

The final step of the experimental procedure is the evaluation phase. All the trained models are evaluated in the same

\begin{figure}[t]
  \centering
  \centerline{\includegraphics[width=9cm,height=5.5cm,keepaspectratio]{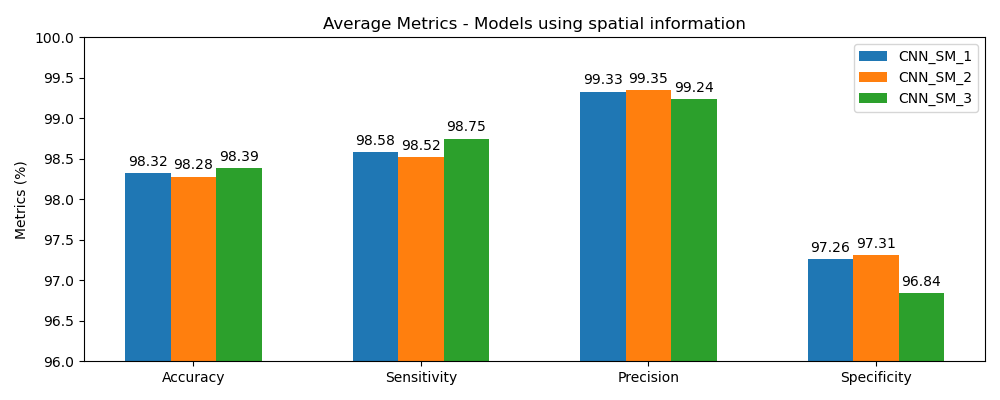}}
  \caption{Evaluation of \(CNN_{sm_1}\), \(CNN_{sm_2}\), and \(CNN_{sm_3}\) models - Average metrics}
  \vspace{-3.5mm}
\end{figure}

\begin{figure}[!ht]
  \centering
  \centerline{\includegraphics[width=9cm,height=5.5cm,keepaspectratio]{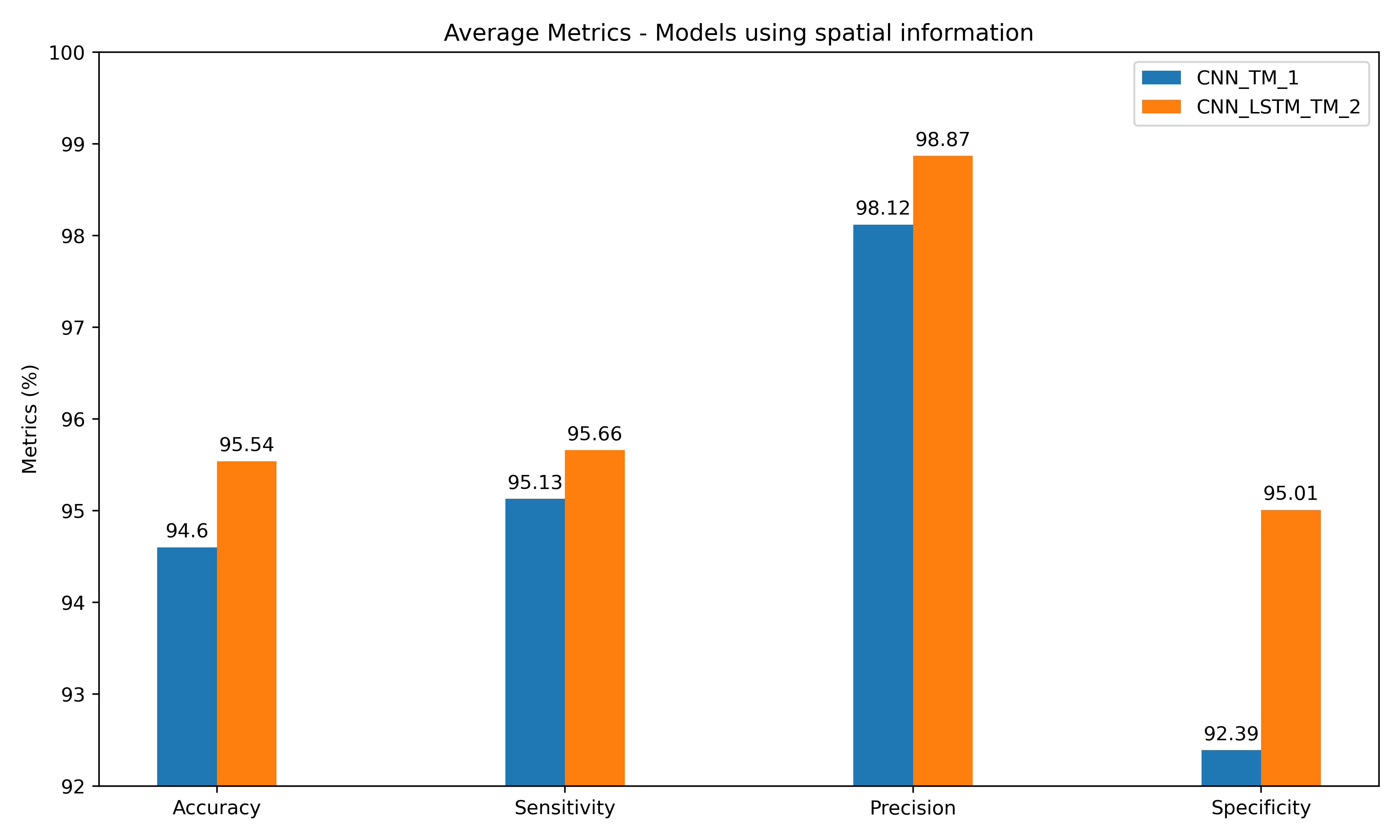}}
  \caption{Evaluation of \(CNN_{tm_1}\) and \(CNN-LSTM_{tm_2}\) models - Average metrics}
  \vspace{-3.5mm}
\end{figure}

\begin{figure}[!ht]
  \centering
  \centerline{\includegraphics[width=9cm,height=5.5cm,keepaspectratio]{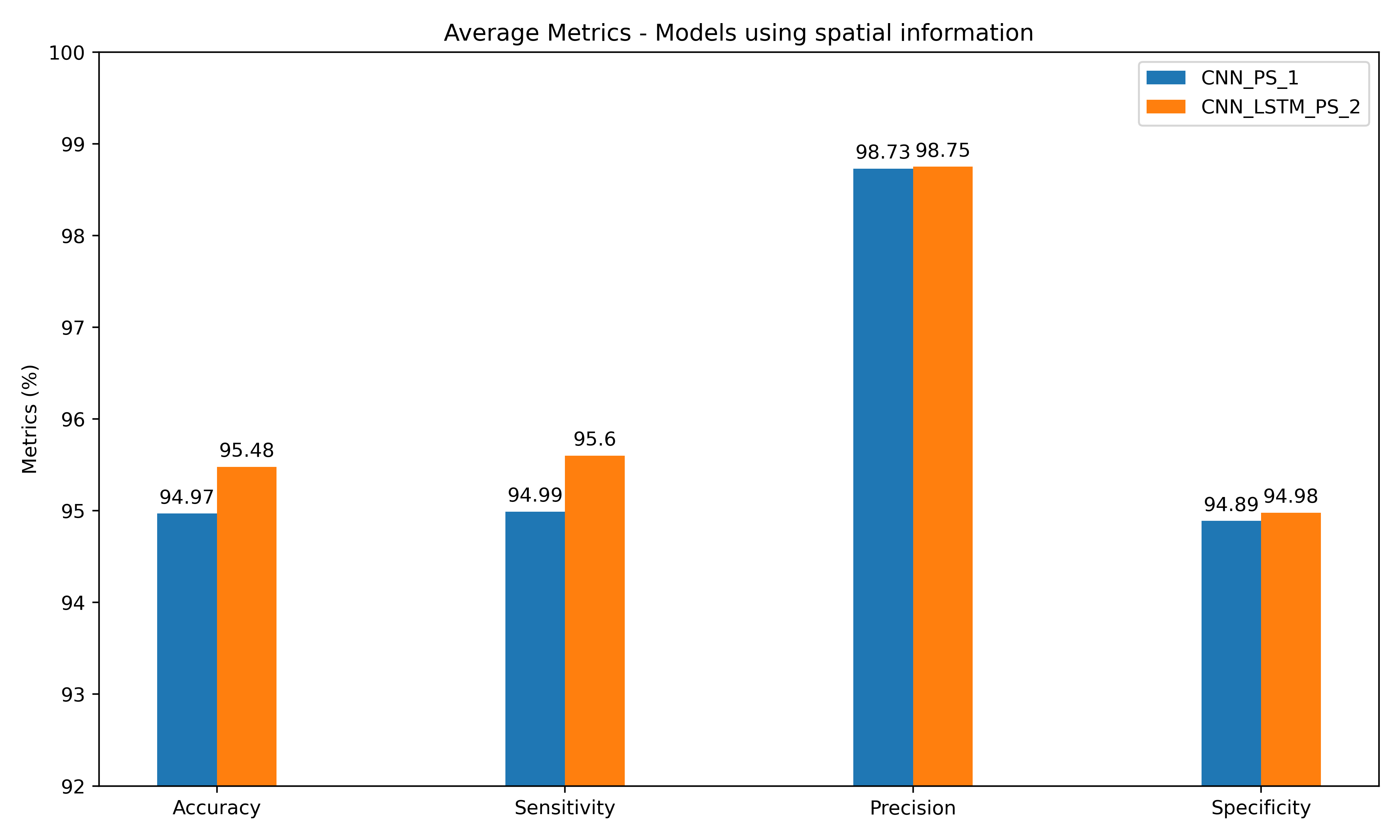}}
  \caption{Evaluation of \(CNN_{ps_1}\) and \(CNN-LSTM_{ps_2}\) models - Average metrics}
  \vspace{-1mm}
\end{figure}

\begin{table}[!ht]
  \centering
  \begin{tabular}{|c|c|c|c|c|} 
    \hline
    \textbf{Model} & {\textbf{ACC}} & {\textbf{SEN}} & {\textbf{PREC}} & {\textbf{SPEC}} \\
    \hline
    \(Comb_{1}\) &95.66 &96 &98.59 &94.29\\
    \hline
    \(Comb_{2}\) &95.62 &95.69 &\textbf{98.85} &\textbf{95.37}\\
    \hline
    \(Comb_{3}\) &95.77 &96.48 &98.26 &92.83\\
    \hline
    \(Comb_{4}\) &\textbf{96.27} &\textbf{96.9} &98.46 &94.67\\
    \hline
  \end{tabular}
  \caption{Evaluation of the combined models - Average metrics (\%)}
  \vspace{-2mm}
  \label{tab:ok}
\end{table}

\noindent test set. Accuracy, precision, sensitivity, and specificity are calculated. The final predictions are extracted separately from each trained model, however different voting schemes using the extracted probabilities are also applied. The models are tested using 294 subjects (test set). For each split (5-fold cross-validation) the four performance metrics (accuracy: ACC, precision: PREC, sensitivity: SEN, and specificity: SPEC) are calculated. Moreover, a voting schema for the final decision is applied in order to evaluate whether combinations of the distinct models result in better performance.

A general description of the weighted voting schemes with $n$ different models is the following: 

\begin{equation}
    Prob_{out} = w_{1}Prob_{1} + ... + w_{n}Prob_{n} \mbox{, \(\sum_{i=1}^{n} w_{i} = 1\)},
\end{equation}

\noindent
where \(w_{i}\) and \(Prob_{i}\) are the voting weight and the extracted probability of the \(i^{th}\) model, respectively. If  \(Prob_{out} > 0.5\) (threshold) then the component is considered classified as an artifact, else it is classified as a neuronal signal. Both the time and frequency information are derived from the same data (time courses of the mixing matrix), hence, we selected the weights in order to balance the contribution of the spatial maps and time courses in the decision function. The evaluated voting schemes (inside the parentheses are the corresponding voting weights) are the following: 

\begin{itemize}
    \item \(Schema_{1}\): \(CNN_{sm_1}\) (0.5), \(CNN_{tm_1}\) (0.25), and \(CNN_{ps_1}\) (0.25)
    \vspace{-2mm}
    \item \(Schema_{2}\): \(CNN_{sm_1}\) (0.5), \(CNN-LSTM_{tm_2}\) (0.25), and \(CNN-LSTM_{ps_2}\) (0.25)
    \vspace{-2mm}
    \item \(Schema_{3}\): \(CNN_{sm_2}\) (0.5), \(CNN-LSTM_{tm_2}\) (0.25), and \(CNN-LSTM_{ps_2}\) (0.25)
    \vspace{-2mm}
    \item \(Schema_{4}\): \(CNN-LSTM_{tm_2}\) (0.5), and \(CNN-LSTM_{ps_2}\) (0.5)
\end{itemize}

\begin{figure}[b]
\vspace{-2.5mm}
  \centering
  \centerline{\includegraphics[width=9cm,height=5.5cm,keepaspectratio]{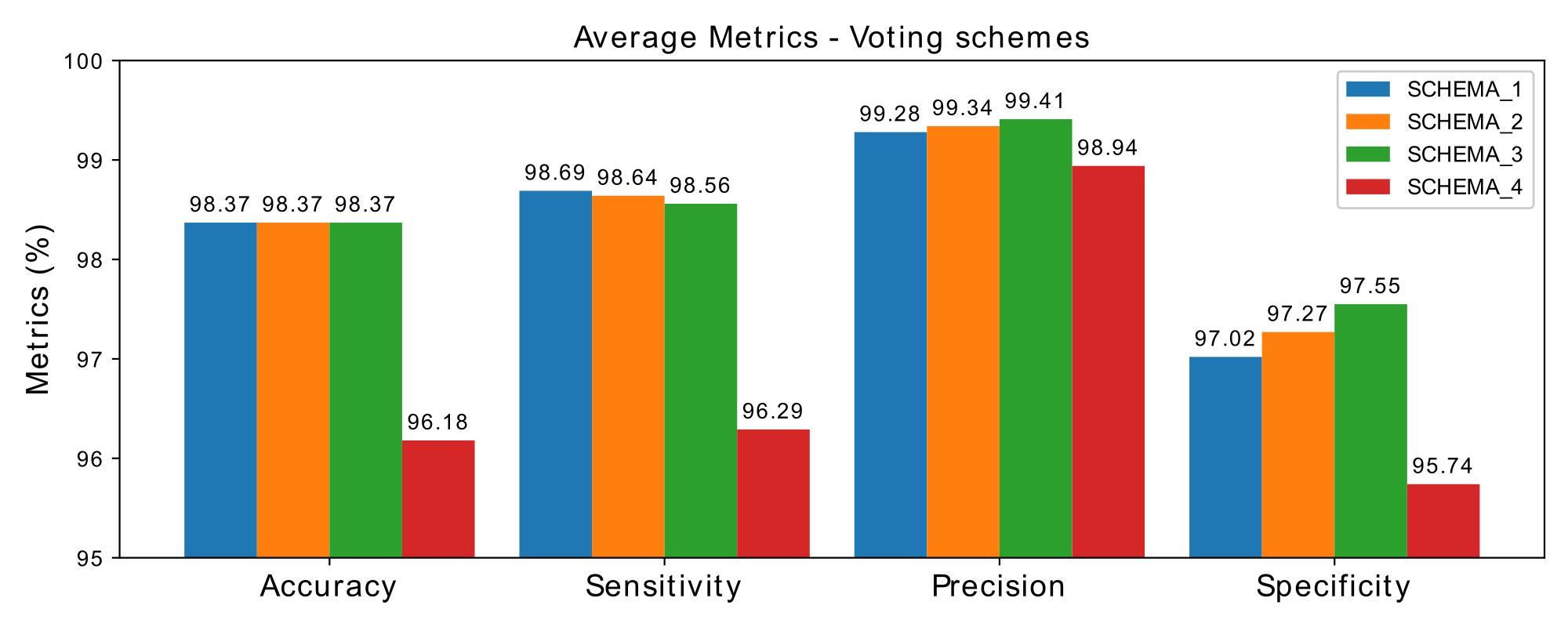}}
  \caption{Evaluation of the voting schemes - Average metrics}
\end{figure}

Figure 6 indicates that the performance of the three different models using spatial information is very similar. The accuracy is over 98\%, so the possible improvement is limited. The addition of the residual blocks (\(CNN_{sm_3}\) model) increases the complexity of the model, but the performance does not improve significantly. Moreover, BN layers which are included in \(CNN_{sm_2}\) model do not affect the performance. 
\par

The models using temporal information (\(CNN_{tm_1}\) and \(CNN-LSTM_{tm_2}\)) perform worse than those using spatial information as the accuracy decreases  approximately by 3\%. The high resolution of the spatial maps is an important aspect of the models' efficiency. Figure 7 shows that the addition of the LSTM block in \(CNN-LSTM_{tm_2}\) model results in better performance as the model is capable of learning better the sequential patterns.  The models using frequency information (\(CNN_{ps_1}\) and \(CNN-LSTM_{ps_2}\)) perform similarly to the models using temporal information. The \(CNN-LSTM_{ps_2}\) model with the LSTM block achieves better performance (Figure 8). 
\par

The evaluation of the combined models \(Comb_{1}\) and \(Comb_{3}\) (Table 1) demonstrates that the end-to-end training using multiple sources of information (spatial, temporal, and frequency) is not advantageous. The \(Comb_{2}\) and \(Comb_{4}\) models perform better than those using one source of information (temporal or frequency). Figure 9 presents the results of the different voting schemes. The performance of the voting schemes 1, 2, and 3 (\(Schema_{1}\), \(Schema_{2}\), and \(Schema_{3}\)) is almost identical. However, \(Schema_{3}\) is slightly more robust and stable as it seems to generalize significantly well using the different splits (5-fold cross validation). 

\section{DISCUSSION AND CONCLUSION}
\label{sec:conclusion}

The results of this study indicate that the denoising and artifact removal of resting-state fMRI can be very effectively implemented using a DNN framework. The models of spatial maps (\(CNN_{sm_1}\), \(CNN_{sm_2}\), and \(CNN_{sm_3}\)) perform almost identically and the accuracy is over 98\%. This finding demonstrates that the usage of high-resolution spatial information, without the addition of temporal information, can present exceptional performance. The temporal models (\(CNN_{tm_1}\), and \(CNN-LSTM_{tm_2}\)) are less efficient than spatial models. It is worth mentioning that the addition of the LSTM block in \(CNN-LSTM_{tm_2}\) model boosts the performance significantly with an accuracy increment of almost 1\% (around 95.5\%). Similarly, the frequency models  (\(CNN_{ps_1}\), and \(CNN-LSTM_{ps_2}\)) perform worse than spatial models and the enhanced model \(CNN-LSTM_{ps_2}\) with the LSTM block achieves higher evaluation metrics compared to \(CNN_{ps_1}\). Notably, the evaluation of combined models (\(Comb_{2}\), and \(Comb_{4}\)) and the voting schema (\(Schema_{4}\)) points out that the combination of timecourses and power spectrum as inputs is valuable and increases the performance (accuracy over 96\%). Hence, the hypothesis that the DNN models learn the features related to frequency automatically, given the temporal information (timecourses), does not entirely hold \cite{kam2019deep}  and adding the frequency information can result in an improved performance of the employed scheme.
\par

The evaluation of the combined models (\(Comb_{1}\), and \(Comb_{3}\)) demonstrates that the joint training using the three channels of information (spatial maps, timecourses, and power spectrum) is not advantageous. Finally, the best results are obtained by the voting \(Schema_{3}\) with average accuracy of 98.37\% and a very good balance between the metrics of sensitivity and specificity. Moreover, this schema shows very stable performance using the different splits in 5-fold cross validation. More precisely, the accuracy is varying from 98.31\% (\(1^{st}\) split) to 98.42\% (\(4^{th}\) split).

\par
The main drawback of the proposed schemes (compared to FIX) is the fact that only healthy adult brains have been used for training the models. Hence, in order to use the proposed scheme in studies with brains of different size or anatomy (e.g. pediatric subjects), we would either need to retrain the selected scheme or use transfer learning. As future work, we intend to explore such cases with transfer learning approaches in order to evaluate the performance of our models in task-related fMRI studies and also in pediatric subjects. 
Furthermore, inception modules \cite{szegedy2015going} can be tested in the DNN models, as they have shown state of the art results in many Deep Learning tasks. In addition, attention mechanisms can be included in the temporal and frequency models.
\vspace{-6mm}

\noindent{\section{ACKNOWLEDGMENTS}
\label{sec:ackn}
\vspace{-1mm}
This research received funding from EIT 19263 – "SeizeIT2: Discreet Personalized Epileptic Seizure Detection Device" and from the Flemish Government (AI Research Program). All authors are affiliated to Leuven.AI - KU Leuven institute for AI, B-3000, Leuven, Belgium.}

\bibliographystyle{IEEEbib}
\bibliography{strings,refs}

\end{document}